\documentstyle[12pt,epsfig]{article}
\def\lsim{\mathrel{\raise3pt\hbox to 8pt{\raise -6pt\hbox{$\sim$}\hss{$<$}}}}
\newcommand{\boldpi}{\mbox{\boldmath $\pi$}}
\newcommand{\boldtau}{\mbox{\boldmath $\tau$}}

\def\haf{\textstyle{1\over2}}
\newcommand{\vecr}{\vec{r}}
\newcommand{\vecsig}{\vec{\sigma}}
\newcommand{\vnabla}{\vec{\nabla}}
\newcommand{\fpi}{f_{\pi}}
\newcommand{\mpi}{m_{\pi}}
\newcommand{\mpipm}{m_{\pi^{\pm}}}
\newcommand{\mpip}{m_{\pi^+}}
\newcommand{\mpim}{m_{\pi^-}}
\newcommand{\mpiz}{m_{\pi^0}}
\newcommand{\bmpi}{\overline{m}_\pi}

\newskip\humongous \humongous=0pt plus 1000pt minus 1000pt
\def\caja{\mathsurround=0pt}

\newif\ifdtup
\def\panorama{\global\dtuptrue \openup1\jot \caja
        \everycr{\noalign{\ifdtup \global\dtupfalse
        \vskip-\lineskiplimit \vskip\normallineskiplimit
        \else \penalty\interdisplaylinepenalty \fi}}}
\def\eqalignno#1{\panorama \tabskip=\humongous
        \halign to\displaywidth{\hfil$\displaystyle{##}$
        \tabskip=0pt&$\displaystyle{{}##}$\hfil
        \tabskip=\humongous&\llap{$##$}\tabskip=0pt
        \crcr#1\crcr}}
\begin{document}
\vspace*{-0.6in}
\hfill \fbox{\parbox[t]{1.53in}{LA-UR-99-1183-REV\\
KRL MAP-251}}\hspace*{0.35in}
\vspace*{0.6in}

\begin{center}

{\Large {\bf Charge-Independence Breaking in the Two-Pion-Exchange
Nucleon-Nucleon Force}}\\

\vspace*{0.20in}
by\\
\vspace*{0.20in}
J.\ L.\ Friar \\
Theoretical Division \\
Los Alamos National Laboratory \\
Los Alamos, NM  87545 \\
\vspace*{0.20in}
and\\
\vspace*{0.20in}
U.\ van Kolck \\
Kellogg Radiation Laboratory, 106-38\\
California Institute of Technology \\
Pasadena, CA 91125 \\
\end{center}
\vspace*{0.25in}

\begin{abstract}
Charge-independence breaking due to the pion-mass difference in the (chiral)
two-pion-exchange nucleon-nucleon force  is investigated. A general argument
based on symmetries is presented that relates the charge-symmetric part of that
force to the proton-proton case. The static potential linear in that mass
difference is worked out as an explicit example by means of Feynman diagrams,
and this confirms the general argument.
\end{abstract}

\pagebreak

Although a complete understanding of isospin violation (IV) in the nuclear force
remains to be achieved, significant progress has been made.  Decades of
experimental progress (reviewed and summarized nicely in
Refs.\cite{iv1,iv2,iv3}) have been supplemented recently by the advent of chiral
perturbation theory (ChPT)\cite{weinberg,iv,texas,bkm}. This powerful technique
casts the symmetries of QCD into effective interactions of the traditional,
low-energy degrees of freedom of nuclear physics (viz., nucleons and pions).
These building blocks (Lagrangians) can then be combined in the usual way to
produce IV nuclear forces.

One of the significant attributes of effective field theories is power counting,
which is used to organize calculations\cite{weinberg,iv,texas,bkm,ndpc,dpc}.
That is, a well-defined ordering of terms in the Lagrangian according to an
intrinsic-size criterion is used to generate all nuclear-force terms of a
particular size.  In addition, IV terms in such theories can be classified
according to whether their origin is the mass differences of the quarks or hard
electromagnetic (EM) interactions at the quark level\cite{iv}. Soft EM
interactions (such as the Coulomb force) can be constructed in the usual
way\cite{pig}.

This scheme was used recently for the first time\cite{iv} to explain the sizes
of the different forms of IV in the nuclear force.  A convenient and
universal\cite{iv1} classification for nuclear isospin is:  class (I) - isospin
conserving; class (II) - charge-independence breaking (CIB) of isospin, but
charge symmetric; class (III) - charge-symmetry breaking (CSB) of isospin; class
(IV) - isospin mixing in the $np$ system between $T=0$ and $T=1$. Power counting
can be used to demonstrate\cite{iv} that class (I) forces are stronger than
class (II), which is stronger than class (III), which is stronger than class
(IV).  Thus, class (II) isospin violation is the largest, and that is the
purview of this work.

Many mechanisms contribute to charge-symmetric CIB, but the largest is due to
the mass difference of the charged and neutral pions,  $\delta \mpi = \mpipm -
\mpiz$, which is primarily of electromagnetic origin.  Indeed most CIB
mechanisms are of this type. The pion-mass difference produces an IV effect of
order $(\delta \mpi /\mpi) \sim 3 \%$ in the one-pion-exchange potential (OPEP).
 Simultaneous pion-photon exchange\cite{pig} is of order $(\alpha/\pi)$ times
OPEP, as would be the effect of EM modification\cite{1loop,mor} of the $\pi$-$N$
coupling constant (not yet detected); both of the latter are class (II)
mechanisms.  In addition there will be short-range CIB forces, of nominal size
$(\alpha / \pi)$ times the usual short-range force; such forces have been
discussed recently\cite{lm} in the context of meson-exchange models of CIB.
Finally, in the two-pion-exchange potential (TPEP) the different pion masses
generate the dominant CIB\cite{hm}, which is also of nominal size $(\alpha/\pi)$
times OPEP.

We present below a general argument for the class (II) isospin violation that is
produced by differing pion masses in the two-pion-exchange nucleon-nucleon
force. The general argument (which applies to arbitrary radial forms, including
the leading order, subleading order, ... in ChPT) will be supplemented by a
specific example, namely the static limit (leading order in ChPT) of that
potential, in order to validate the general argument. We also note that the
general argument can be easily incorporated in partial-wave analyses such as 
that carried out by the Nijmegen group in their analysis of nucleon-nucleon
scattering\cite{rob_mart}. Finally, we estimate the effect of the leading-order
IV on the $^1\!S_0$ scattering lengths.

\begin{figure}[htb]
\epsfig{file=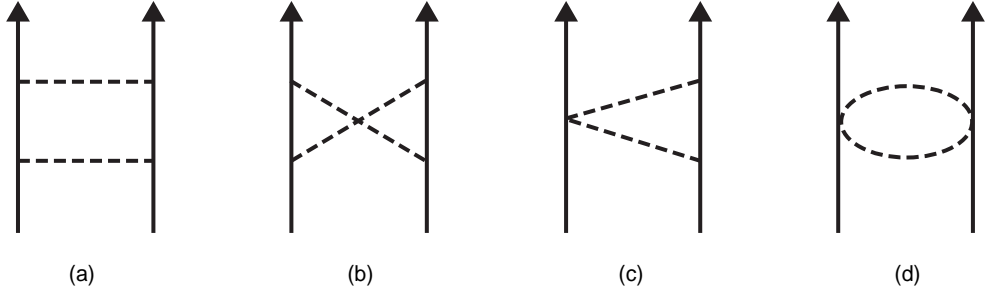,height=1.5in}
\caption{Two-pion-exchange graphs for nucleon-nucleon scattering.}
\end{figure}

Two-pion-exchange isospin-conserving forces are an old problem with a new twist.
In static order (containing only terms that remain when the nucleon mass, $M$,
or the large-mass scale of QCD, $\Lambda$, becomes very large) the diagrams of
Fig. (1) contribute to the TPEP. The vertices and propagators follow from the
leading-order Lagrangian for pions and nucleons,
$$
 {\cal L}^{(0)}  = \frac{1}{2}[\dot{\boldpi}^{2}-(\vec{\nabla}\boldpi)^{2}
          -\mpi^{2}\boldpi^{2}] 
   + N^{\dagger}[i\partial_{0}-\frac{1}{4 \fpi^{2}} \boldtau \cdot
         (\boldpi\times\dot{\boldpi})]N +\frac{g_{A}}{2 \fpi} 
 N^{\dagger}\vecsig \cdot\vec{\nabla}(\boldtau\cdot\boldpi)N \, , \eqno (1)
$$
where the $\pi \pi N$ term is the Weinberg-Tomozawa (WT) interaction\cite{wt}
and the $\pi N$ term is the usual interaction that depends on the axial-vector
coupling constant, $g_A (\simeq 1.25)$, and the pion-decay constant, $\fpi
(\simeq 92$ MeV).  Terms with additional pions or nucleons are neglected here,
as they only contribute to the nuclear force at higher orders. The WT term has a
specific normalization ($-1 / 4 \fpi^2$) required by the underlying chiral
symmetry.

The first treatments of the box (Fig. (1a)) and crossed-box (Fig. (1b)) diagrams
that led to an energy-independent potential were performed by Taketani, Machida,
and Ohnuma (TMO)\cite{tmo}, and by Sugawara and Okubo (S-O)\cite{s-o}.
Phenomenological Lagrangians of undetermined size were also incorporated by S-O,
including a term of WT type. The first calculation based on a chiral Lagrangian
(the new twist) was performed in Ref.\cite{texas}, and this has been repeated by
several groups\cite{fc,munich,evgeni}. The result (short-range terms have been 
ignored) is conveniently expressed in terms of isospin factors as
$$
V_{2 \pi} = V_{2 \pi}^0 + V_{2 \pi}^1 \boldtau_1 \cdot \boldtau_2 \, , 
\eqno (2a)
$$
$$\eqalignno{
V_{2 \pi}^0 = -\frac{\mpi}{8 \pi^3} \left(\frac{g_A \mpi}{2 \fpi}\right)^{\! 4}
&\left ( S_{12} \left[\frac{12 K_0 (2x)}{x^3} + \frac{K_1 (2x)}{x^2}
(4 + \frac{15}{x^2})\right] \right. &{ }\cr
&\left. -4 \, \vecsig_1 \cdot \vecsig_2 \left[\frac{3 K_0 (2x)}{x^3} + 
\frac{K_1 (2x)}{x^2} (2 + \frac{3}{x^2})\right] \right ) \, , &(2b) \cr}
$$
$$\eqalignno{
V_{2 \pi}^1 = \frac{\mpi}{4 \pi^3} \left (\frac{g_A \mpi}{2 \fpi}\right )^{\! 4}
& \left ( -\frac{1}{2}
\left [\frac{ K_0 (2x)}{x}(4 + \frac{23}{x^2}) + \frac{K_1 (2x)}{x^2}(12 +
\frac{23}{x^2})\right ] \right. &{ }\cr
&\left. + \frac{2}{g_A^2} \left [\frac{ K_1 (2x)}{x^2} + 
\frac{5 K_2 (2x)}{2 x^3} \right ] +\frac{K_2 (2x)}{2 g_A^4 x^3} \right ) 
\, , &(2c) \cr}
$$
where $x = \mpi |\vecr_1 - \vecr_2| =\mpi r$, $\vecsig_i$ and $\boldtau_i$ are
the usual (Pauli) spin and isospin operators of nucleon ``$i$'', while $S_{12}$
is the conventional tensor operator and the $K_n$ are irregular Bessel
functions. Terms proportional to $g^4_A$ arise from the box and crossed-box
diagrams (Figs. (1a) and (1b)) with the iterated OPEP appropriately subtracted,
$g^2_A$ terms from the triangle diagrams (Fig. (1c)), and the $g_A$-independent
terms from the ``football'' diagram (Fig. (1d)) constructed from two WT
interactions.

The chiral Lagrangian also contains interactions with more derivatives or powers
of the quark masses, which include new seagull vertices giving rise to a
subleading TPEP of the triangle type. Such forces have also been
calculated\cite{texas,munich,la}, and some components (such as the central
isoscalar) are known to be important\cite{munich,nijmegen}. A new Nijmegen
phase-shift analysis of $pp$ data\cite{nijmegen} has shown that this TPEP
(leading plus subleading order) provides a better fit than the OPEP alone, and
even better than the OPEP supplemented by heavier-meson exchange. In
anticipation of a re-analysis of $np$ data, we examine here CIB in the TPEP.

The pion-mass difference corresponds to a Lagrangian term that arises primarily
from EM interactions,
$$
{\cal L}_{\rm EM}^{(1)} = -\frac{\delta \mpi^2}{2} (\boldpi^2 -\pi^2_3 ) \, . 
\eqno (3)
$$
The superscript $(1)$ here reflects the expected size of this term compared with
the ($\pi^0$) mass in Eq.~(1): on the basis of dimensional
analysis\cite{ndpc,dpc}, $\delta \mpi^2/m_\pi^2 \sim (\alpha/\pi)
(\Lambda/m_\pi)^2$, which numerically is $\sim (m_\pi/\Lambda)$\cite{iv}.
Combining the mass terms in Eqs.~(1) and (3) produces different masses for the
$\pi^{\pm}$ and the $\pi^0$ and correspondingly different pion propagators.
Isospin violation arising from differing pion masses can then be implemented in
a straightforward way by tagging pion masses in the propagators of Fig. (1) with
the isospin labels at the vertices that created them. A straightforward (but
tedious) calculation using the techniques of Ref.\cite{fc} leads to the results
shown below. In the sense of power counting, these contributions are of the same
order as the isospin-symmetric, subleading TPEP already incorporated in the new
Nijmegen phase shift analysis\cite{nijmegen}. A much simpler derivation (than 
using tagged pion masses) is possible that is based on symmetries, and that is 
what is presented next. This derivation subsumes our leading-order result and
applies to subleading orders, as well.

The defining aspect of the problem is the equality of $\pi^+$ and $\pi^-$
masses, which follows from CPT invariance, and this equality has been
incorporated into Eq.~(3).  Under the reflection in the $x$-$z$ isospin plane
that defines the charge-symmetry operation ($x \rightarrow -x \,; z \rightarrow
-z$), the interaction in Eq.~(3) is invariant and therefore can generate only
class (II) isospin violation, which is charge symmetric. For two nucleons
(labeled 1 and 2) there is a unique isospin operator with this structure,
$$
T_{20} (1,2) = \tau_1^z \tau_2^z - \frac{\boldtau_1 \cdot
\boldtau_2}{3}\, . \eqno (4)
$$
This is an isotensor with vanishing value for $T=0$ states, while for $T=1$
systems it has equal values for $pp$ and $nn$ channels $(2/3)$ and a different
value $(-4/3)$ for the $np$ channel. We write a class (II) potential as
$$
V_{\rm II}=  T_{20} (1,2)\, \Delta V^{\rm CIB} \, .
\eqno (5)
$$

To illustrate our method, let us consider first the well-known effect of the
pion-mass difference in the OPEP\cite{hm}. It follows from the fact that there
are two charged and one neutral pion. If one expands the exact OPEP to first
order in the pion-mass difference,  $\delta \mpi$, about an average pion mass
$$
\bmpi = \frac{2}{3}\mpip + \frac{1}{3}\mpiz \,  , \eqno (6)
$$
one easily finds an isospin-symmetric, isovector $V_\pi^{1}(\bmpi)$ plus the CIB
piece
$$
\Delta V_\pi^{CIB} = -\left (\frac{g_A}{2\fpi}\right )^{\! 2} 
\frac{\delta \mpi}{4\pi}\, \vecsig_1 \cdot \vnabla \,
\vecsig_2 \cdot \vnabla \, e^{-\bmpi r}\,  . \eqno (7)
$$
Alternatively, Eq.~(7) follows from the $pp$ or $nn$ results (where charge
conservation requires that the exchanged pion be neutral) by expanding
$\pi^0$-exchange about $\bmpi$. Equation~(6) is a very commonly used
prescription in nuclear force models.

The TPEP isospin structure is simplest for the $pp$ or $nn$ cases, where charge
conservation requires that the pair of exchanged pions be either both neutral or
both charged (i.e., $\pi^+ \pi^-$). The isoscalar potential $V_{2\pi}^0$
(Eq.~(2b)) must arise from a trace, or a sum over charge states (2 charged pions
for each neutral one). Thus, the actual form of $V_{2\pi}^0$ is $[\frac{2}{3}
V_{2 \pi}^0 (\mpip ; \mpim) + \frac{1}{3} V_{2\pi}^0 (\mpiz ; \mpiz)]$ (where
the masses of both exchanged pions have been explicitly labeled using an obvious
notation), since there are two charged-pion pairs for each neutral pair. This
exact result can also be expanded to first order in $\delta \mpi$ about $\bmpi$.
The expression $V_{2\pi}^0 (\bmpi ; \bmpi)$ then approximately equals the
expanded expression (or equivalently that the term proportional to $\delta \mpi$
cancels out). Thus, using $\bmpi$ in $V^0_{2 \pi}$ is correct through ${\cal
O}(\delta \mpi)$.

The isospin-dependent terms can be deduced by examining the isospin structure of
the WT interaction for emitting two pions with isospins $\alpha$ and $\beta$:
$\tau^\gamma \varepsilon^{\gamma \alpha \beta}$. The $pp$ (or $nn$) interaction
requires $\gamma = 3$, and therefore $\alpha$ and $\beta$ must be $1$ and $2$
(in either order), requiring that both exchanged pions must be charged. The box
diagrams (and also diagrams involving any seagulls, including higher-order ones)
have similar structures. In general, two pions emitted sequentially on a single
nucleon line have an isospin factor $\tau^\alpha \tau^\beta$ (for pions with
isospin $\alpha$ and $\beta$) that can be decomposed into two irreducible terms:
 $\delta^{\alpha \beta}$ and $\varepsilon^{\alpha \beta \gamma} \tau^\gamma$. 
When contracted with the corresponding factors on the second nucleon, the
$\delta^{\alpha \beta}$ leads to the isoscalar force discussed above and the
second factor generates the isovector one proportional to $\tau_1^z \tau_2^z$. 
These are also the allowed structures for seagulls.  Thus, our argument applies
to any $2\pi$-exchange force of this type.

Summarizing, only a pair of charged pions can be exchanged in the isovector
$(\boldtau_1 \cdot \boldtau_2)$ part of the force appropriate for two protons or
two neutrons.  Incorporating the $np$ case, the structure turns out to be
$$\eqalignno{
V_{2 \pi}^1 &\sim \tau_1^z \tau_2^z \, V_{2\pi}^1 (\mpip ; \mpip) + 
(\boldtau_1 \cdot \boldtau_2
- \tau_1^z \tau_2^z) V_{2\pi}^1 (\mpip; \mpiz) \cr
&\rightarrow 2\, V_{2 \pi}^1 
(\mpip ; \mpiz) - V_{2 \pi}^1 (\mpip;\mpip) \, , &(8)\cr}
$$
where the second term on the first line contributes only to $np$ interactions,
and the second form applies only to the $T=1$ $np$ case. The integral
corresponding to $V_{2 \pi}^1 (\mpip;\mpiz)$ is intractable. However, if one
expands the second form about $\bmpi$ one finds that to ${\cal O}(\delta \mpi)$
it is equivalent to $V_{2 \pi}^1 (\mpiz;\mpiz)$, which is a standard integral.
Thus, exchanging two neutral pions in the isovector part of the force is an
excellent approximation for the $T=1$ $np$ case. For $T=0$ there is no CIB and
Eq.~(8) (top line) leads to $V_{2 \pi}^1 (\bmpi;\bmpi)$ as an excellent
approximation.

In the $pp$ case $(\tau_1^z \tau_2^z \rightarrow 1$ and $T_{20} (1,2)
\rightarrow 2/3)$ if the pion masses are expanded about $\bmpi$, the
isospin-violating force can be immediately deduced, since the part of $V_{2
\pi}^1$ proportional to $\delta \mpi$ is isospin violating
$$
V_{2 \pi}^1 (\mpip ; \mpip) \cong V_{2\pi}^1 (\bmpi ; 
\bmpi) + \frac{\delta \mpi}{3} \frac {\partial \; \; \; \;}{\partial 
\bmpi} V_{2 \pi}^1 (\bmpi ; \bmpi) \,  , \eqno (9)
$$
where the first term is isospin conserving.  Because of the $2/3$ value of
$T_{20}$, we arrive at the simple result
$$
\Delta V_{2\pi}^{CIB}=\frac {\delta \mpi}{2} 
               \frac {\partial\; \; \; \; }{\partial \bmpi} 
               V_{2\pi}^1 (\bmpi ; \bmpi) . \eqno (10)
$$
Incorporating the $np$ interaction leads to the same result, which was verified
by following the isospin of each exchanged pion in Fig.~(1) in a conventional
derivation\cite{fc}.

Using the TMO potential in Eq.~(2c) and performing the derivative in Eq.~(10)
one finds
$$\eqalignno{
\Delta V_{2\pi}^{CIB} = 
- \frac{\delta \mpi}{4 \pi^3} \left (\frac{g_A \mpi}{2 \fpi}\right )^{\! 4}
& \left (-\frac{1}{2}\left [\frac{4 K_0 (2x)}{x}+ 
K_1 (2x)(4 + \frac{11}{x^2})\right ] \right. &{ }\cr
&\left. +\frac{3 K_1 (2x) + 2 x\, K_0 (2x)}{g_A^2 x^2}
+\frac{K_1 (2x)}{2 g_A^4 x^2} \right ) \, . &(11)\cr}
$$
Terms proportional to $g_A^4$, $g_A^2$, and $g_A^0$ are again box, triangle, and
football contributions. Equation~(10), which leads to Eq.~(11) in leading order
of ChPT, is our primary result. To ${\cal O}(\delta \mpi)$ Eq.~(11) is identical
to that of Ref.\cite{hm} for the box diagrams, which were also calculated using
the TMO approach. We note that our mass-expansion technique is similar to that
of Ref.\cite{cs}.

Although these forces are appropriate for the Nijmegen phase-shift analysis,
which uses only the tail of the force, a complete potential must be regularized
by a short-range cut off.  We have employed the representations of the appendix
of Ref.\cite{fc} in an attempt to generate analytic forms for the regularized
potential, but have failed. Presumably either purely numerical forms of $\Delta
V_{2 \pi}^{CIB}$ must be generated or approximations made to simplify
intractable integrals.

We have resorted instead to an {\it ad hoc} cutoff procedure. We calculate the
CIB effect using the charge-independent part of the AV18 potential\cite{av18}
with the electromagnetic corrections turned off, and we use that potential's
multiplicative two-pion-range cutoff: $(1-e^{-2.1\, r^2})^2$. We find that the
separate contributions of the [box,triangle,football] potentials to $\Delta a =
|a_{np}| -\haf |a_{pp}+a_{nn}|$ are: [0.98,-0.31,-0.02]~fm for a total of
0.65~fm. The triangle contribution from the WT Lagrangian is a sizable
correction to the dominant box graphs. We are aware of only two previous
comparable calculations of the CIB from the pion-mass difference in
two-pion-exchange forces, and neither can be directly compared to our result. Li
and Machleidt\cite{lm} find 0.16~fm from the box graphs, but their calculation
includes the (cutoff) delta functions that we have eliminated by
renormalization. Ericson and Miller\cite{em} find 0.88~fm using the relativistic
PS-coupling model of Partovi and Lomon\cite{pl}. Although the latter is in
reasonable agreement with our results, differences in the two approaches may
make this agreement accidental.

In summary, for the first time the leading-order (static) chiral CIB $NN$ force
from $2\pi$ exchange has been developed, employing both symmetry arguments and
direct calculation of Feynman diagrams. The symmetry arguments apply only to the
CIB from the pion-mass difference, but are appropriate to any order in ChPT. We
find that to ${\cal O}(\delta \mpi)$ the effective pion mass to be used in the
isoscalar force or the $T=0$ force is $\bmpi$, while $\mpipm$ should be used for
the $pp$ or $nn$ cases and $\mpiz$ for the $T=1$ $np$ case.

\begin{center}
{\large {\bf Acknowledgments}}\\
\end{center}
The work of JLF was performed under the auspices of the United States Department
of Energy, while that of UvK was supported under National Science Foundation
grant PHY 94-20470. We would like to thank R. Timmermans of KVI, E. M. Henley of
the University of Washington, and S. A. Coon of New Mexico State University for
their helpful comments and discussion.


\begin{thebibliography}{999}
\bibitem{la} J. L. Friar, {\tt nucl-th/9901082}. This paper calculates the 
        subleading-order TPEP for a variety of off-shell extensions of OPEP.
        The forces due to the dominant counter-term seagulls are not affected by
        such considerations.
\end{thebibliography}
\end{document}